\documentclass[preprint,superbib,floats]{revtex4}
\pdfoutput=1
\usepackage{graphicx,color}
\usepackage[centertags]{amsmath}
\usepackage{paralist}
\usepackage{tabularx}
\usepackage{delarray}
\usepackage{bm}

\begin{document}
\title{Ultrafast single electron spin manipulation in 2D semiconductor quantum dots with optimally controlled time-dependent electric fields through spin-orbit coupling}

\author{J. A. Budagosky}
\affiliation{Institute for Biocomputation and Physics of Complex Systems,
  University of Zaragoza, Mariano Esquillor s/n, 50018 Zaragoza (Spain)}

\author{A. Castro}
\affiliation{ARAID Foundation, Edificion CEEI, Mar{\'{\i}}a Luna 11, 50018 Zaragoza (Spain)}
\affiliation{Institute for Biocomputation and Physics of Complex Systems,
  University of Zaragoza, Mariano Esquillor s/n, 50018 Zaragoza (Spain)}

\date{\today}
\begin{abstract}
  We have studied theoretically the possibility of ultra-fast manipulation of
  a single electron spin in 2D semiconductor quantum dots, by means of
  high-frequency time-dependent electric fields. The electron spin degree of
  freedom is excited through spin-orbit coupling, and the procedure may be
  enhanced by the presence of a static magnetic field. We use quantum optimal
  control theory to tailor the temporal profile of the electric field in order
  to achieve the most effective manipulation. The scheme predicts significant
  control over spin operations in times of the order of picoseconds -- an
  ultrafast time scale that permits to avoid the effects of decoherence if
  this scheme is to be used as a tool for quantum information processing.
\end{abstract}
\maketitle

\section{Introduction}

Quantum dots are nanoscopic artificial structures, somehow created within
solid state devices, that contain a small number of charge carriers (electrons
or holes), and display quantum behaviour in the same manner that atoms or
molecules do~\cite{Kouwenhoven2001}.  Nowadays it is relatively easy to
isolate even one single electron.  The precise control over charge and
currents on quantum dots was soon achieved and demonstrated, whereas the
experimental techniques to measure and manipulate the spins followed later
on~\cite{Hanson2007}.  These developments have contributed to the growth of
the field of \emph{spintronics}~\cite{Zutic2004}, which consists of the
control and manipulation of the spin degrees of freedom in solid state
devices.

Among the many foreseen applications of quantum dot based spintronics is
quantum information processing: the spin of a single electron is the most
typical example of a two-level system, and this fact soon suggested the
possibility of using single-electron quantum dots as a physical realization
of a \emph{qubit}~\cite{Loss1998}. One of the reasons for this to be
conceivable is the long decoherence times observed for the spin degree of
freedom in common quantum dots~\cite{Elzerman2004-2,Taylor2007} (other reason
is the realization, at least at the level of proof-of-principle, of the
DeVincenzo's criteria~\cite{DeVincenzo2000}). This decoherence time is
to be compared with the time required for an \emph{operation}, i.e. the
typical time used to change the state of the system in a controlled manner,
with an external field. It is therefore essential to have a means to produce
very fast operation times, specially since fault-tolerant operations require
multiple possible manipulations within the coherence time.

In order to induce a spin flip in these systems, one can of course use a
time-dependent magnetic field, oscillating at the Zeeman transition frequency;
the Rabi oscillations will eventually induce full transitions from one state
to another. This procedure is called ``electron spin resonance''
(ESR). However, this method is not particularly fast, and moreover it is not
easy to produce and localize these magnetic fields individually on each
quantum dot. Recently, the alternative use of electric fields has also been
proposed and
demonstrated~\cite{KCNowak2007,Stano2008,Press2008,YBan2012,deGreve2013},
since these can indirectly couple to the spin, for example through spin-orbit
coupling. An electric field may be produced locally on chip through
appropriate gates, or one may attempt optical manipulation via a THz laser
pulse, which has the advantage of a very high frequency, and therefore promise
very fast spin rotations.

In this work we will focus on this second option, the optical
manipulation of spins -- or, in any case, the use of high frequency
electric fields, whatever its origin. The action of THz optical fields
on quantum dots, and its coupling to the spin degree of freedom
thorough spin-orbit coupling, has already been theoretically
investigated for example by Jiang et al.~\cite{Jiang2006} In this
work, we inquire into how fast can a single spin in a quantum dot
rotate when manipulated with a THz electric field, through the
indirect coupling facilitated by the spin-orbit interaction. We will
not limit the allowed external fields to quasi-mono-chromatic laser
pulses (that are typically tuned to some resonant frequency), but
rather consider the possibility of shaping the temporal shape of the
pulse, in order to find a path that produces the transition in a
faster way than following Rabi's oscillations.  In order to find the
shape that produces the optimal result, we will use quantum optimal
control theory (QOCT)~\cite{Brif2010,Werschnik2007}. This theory
provides a set of techniques to find the best external fields that,
acting on a quantum system, produce the evolution that is optimal in a
certain sense defined by the user.

We have implemented QOCT in the real-space, real-time, electronic structure
code {\tt octopus}~\cite{Marques2003,ACastro2006,Andrade2012}. This code
focuses on the time-dependent many-electron problem, based on time-dependent
density-functional theory~\cite{Runge1984,Marques2012}, although in this case
this feature is not necessary since we deal with a single-electron
problem. This implementation of QOCT has already been employed in recent years
to study dynamics of 2D quantum dots and rings in the presence of THz laser
fields~\cite{Rasanen2007,Rasanen2008,Rasanen2008-2}. By including in the model
a spin-orbit coupling term (in this case, we have chosen the Rashba
term~\cite{Bychkov1984}, although this choice is not important for the
conclusions that follow), we have learned what are the typical transition
velocities that one may expect when manipulating the spin of a single electron
quantum dot with this kind of electric fields.

The goal has been to construct optimal laser pulses that drive a
single-electron spin from a given initial orientation in the Bloch
sphere to any other \textit{target} spin orientation. Within QOCT,
such physical goals have to be mathematically formalized with the
definition of a ``target function'' that determines the degree of
success achieved for the task that is pursued. In this work we have
considered two possible target definitions: The first one corresponds
to the projection into some pre-defined spin orientation: the control
is exerted on the orientation of spin without specifying a priori
which stationary states are involved. The second target corresponds to
the transition from the ground state, which is known to have a
dominant spin in one particular direction, to an excited eigenstate
that has a dominant spin in an approximately opposite direction.

\section{Theory}

We consider the quantum dots fabricated on top of the two-dimensional electron
gas (2DEG) that can be locked at the interface of a semiconductor
heterostructure. The most common case is perhaps that of the GaAs/AlGaAs
heterostructure, and therefore we will consider this material in the
following. The AlGaAs layer is usually doped with Si, which results in the
liberation of free electrons, that accumulate at the GaAs/AlGaAs interface,
and are trapped in a thin (around 10~nm) layer. The electronic system can then
be considered to occupy a thin potential well in the direction perpendicular
to the interface plane (hereafter, the $z$-direction). This thinness (to be 
compared to the Fermi wave length of the electrons, large due to the low
electronic density) is the reason for the 2D character of the system, as the
electrons can be considered to occupy only the lowest subband -- at the low
temperature that are necessary for these experiments to take place. Once the
2DEG is thus created, one may further constrain electrons in the $xy$ plane,
by placing \emph{gates} (metal electrodes) on top of the semiconductor, and
controlling their voltages.

The electronic islands created in this manner (the \emph{quantum dots}) can
then be modeled by considering an effective mass approximation in 2D, and
assuming simple and smooth confining potentials in the $xy$ plane - typically,
as we will do below, of parabolic form. In addition, one may have an external
magnetic field, a time-dependent external electric field (originated by the
variation of the potential applied on the gates, or by a laser source), and,
as we will crucially consider in this work, one or more spin-orbit coupling
(SOC) terms.

There are various forms of SOC that can be present in this kind of zincblende
materials. The bulk crystal lacks inversion symmetry, which leads to the
Dresselhaus term~\cite{Dresselhaus1955}, and in addition the heterojunction
produces a structural inversion asymmetry along the growth direction that
results in the Rashba term~\cite{Bychkov1984}. The strength of the Rashba
effect can in fact be tailored with the application of external electric
fields applied in parallel to the growth direction. Because of this tunability,
we have chosen to work exclusively with this Rashba term, which is in many circumstances the dominant
one~\cite{Pfeffer1995,Jusserand1995}. However, the qualitative conclusions
that we will draw out do not depend on this choice: these SOC terms couple to
the external electric fields in a similar manner.

After all these considerations, the system can be modeled, in the absence of
external time-dependent electric pulses, with the following static effective Hamiltonian:
\begin{equation}\label{eq-1}
\hat{H}_0=\frac{\hbar^2}{2m^*}\left(-i\bm{\nabla}-\frac{e}{\hbar}\bm{A}(\hat{\bm r})\right)^2 + V_c(\hat{\bm{r}})
+ \alpha\left[\hat{\bm{\sigma}}\times\left(-i\bm{\nabla}-\frac{e}{\hbar}\bm{A}(\hat{\bm{r}})\right)\right]_z + \frac{g^*}{2}\mu_B \left(\hat{\bm{\sigma}}\times\bm{B}\right)_z~~\text{.}
\end{equation}
The first term corresponds to the electron kinetic energy, where $m^*$ is the
electron effective mass that we consider to be $m^*=0.067m_e$ in a GaAs
semiconductor medium ($m_e$ is the electron mass). The vector potential $\bm
{A}(\hat{\bm{r}})$ included in that term generates the static homogeneous magnetic field
$\bm{B}=B\bm{z}$, normal to the $xy$-plane where the system is confined. The second term, 
\begin{equation}
V_c(\hat{\bm{r}})=\frac{1}{2}m^*\hbar\omega_0\left(\hat{x}^2+\hat{y}^2\right)
\end{equation}
is the confinement potential. The third term
correspond to the Rashba SOC, whereas the fourth is the Zeeman term. In
those expressions,
$\hat{\bm{\sigma}}=\left(\hat{\sigma}_x,\hat{\sigma}_y,\hat{\sigma}_z\right)$ is the
vector of Pauli matrices, $\alpha$ is the Rashba parameter that
determines the SOC strength, $g^*$ is the effective gyromagnetic factor (we
will use $g^*=-0.44$ for GaAs), and $\mu_B$ is the Bohr magneton.

For the confinement potential, we have used $\hbar\omega_0=1.8$~meV, which is
in the range of the typical values in common experimental realizations of QDs.
A rough corresponding estimate of the QD lateral extension is approximately
$2\sqrt{\hbar/m^*\omega_0}\approx50$~nm, which also lies within the typical
range of sizes for lithographically etched and gate-confined QDs. In the
following, we will express all the quantities in \emph{effective atomic
  units}, that relate to 
usual atomic units (defined by setting $e^2 = m_e = \hbar = 1$) in the
following manner: $a_0^*=a_0\left(m^*/\kappa\right)$,
$H_a^*=H_a\left(\kappa^2/m^*\right)$, $t_0^*=t_0\left(m^*/\kappa^2\right)$,
where $a_0$, $H_a$ and $t_0$ are the usual atomic units of length, energy, and
time, respectively. The value of $\kappa$ is $13.18~\epsilon_0$ for GaAs.

In the presence of an external electric pulse, the previous Hamiltonian must be
supplemented with a time-dependent term, and the system is governed by the time-dependent Schr\"{o}dinger's equation during a time interval [0, T]:
\begin{equation}\label{eq-2}
i\frac{\partial}{\partial t}\Psi(\bm{r},t)=\hat{H}(t)\Psi(\bm{r},t)=\left[\hat{H}_0-\bm{\hat{\mu}}\bm{\epsilon}(t)\right]\Psi(\bm{r},t)\,,
\end{equation}
where the electron-field interaction assumes the dipole approximation
in the length gauge, being $\bm{\hat{\mu}}=-e\hat{\bm{r}}$ the dipole operator. The
time-dependent electric field,
$\bm{\epsilon}(t)=\epsilon(t)\bm{\pi}$ is linearly polarized in some
direction determined by the unit vector $\bm{\pi}$, contained in the
$xy$-plane. The precise direction is in fact irrelevant due to the circular
symmetry of the rest of the Hamiltonian. 

The specification of $\epsilon(t)$, together with an initial value condition,
determines the full evolution of the system, via the propagation of
Schr\"{o}dinger's equation. The
questions that we wish to answer are the following: is it possible, by
fine-tuning the form of $\epsilon(t)$, to manipulate at will the spin state of
the system in a controllable manner? How fast can this manipulation be
performed, as a function of the characteristics of the system -- the SOC
strength, the presence and magnitude of the external static magnetic field,
etc.?

Optimal control is a suitable tool to address this type of questions,
reformulating them into the following problem: given a \emph{target}, defined
as the maximization of a functional of the final state of the system (or of
its evolution), what is the time-dependent control function that best
accomplishes it? In our case, the target must obviously be related to the
spin state of the system, whereas the control function is the time-dependent
electric field $\epsilon(t)$. The set of possible control functions is the
search space for the optimization algorithm.
In practice, the control function must be discretized in order to proceed with
the numerical computations: a set of parameters $u_1,\dots,u_M\equiv {\bf u}$
determines the shape of the function: $\epsilon(t)=\epsilon[{\bf u}](t)$, and
therefore the domain of the parameters ${\bf u}$ is the effective search space.

Regarding the target, it is typically defined through the expectation value of some
operator $\hat{O}$, i.e. it is a functional of the final state of the system
with the form:
\begin{equation}
F[\Psi] = \langle \Psi(T)\vert\hat{O}\vert\Psi(T)\rangle\,.
\end{equation}
Since the parameters ${\bf u}$ determine the shape of the control
function, which in turns determines the evolution of the system, ${\bf
  u} \to \Psi[{\bf u}]$, the problem is reduced to the maximization of
a function of ${\bf u}$:
\begin{equation}
G[{\bf u}] = F[\Psi[{\bf u}]] = \langle \Psi[{\bf
    u}](T)\vert\hat{O}\vert\Psi[{\bf u}](T)\rangle\,.
\end{equation}
This maximization is greatly eased if we have a feasible scheme to compute the
gradient of this function, and this is provided by QOCT:
\begin{equation}\label{eq-4}
\nabla_{\bf u}G\left[{\bf u}\right] = 
2\text{Im}\;\int_0^T\!\!{\rm d}t\;\langle\chi\left[{\bf u}\right](t)\vert\;\nabla_{\bf
  u}\hat{H}[{\bf u}](t)\;\vert \Psi\left[{\bf u}\right](t)\rangle\,, 
\end{equation}
Note that, given the structure of our Hamiltonian:
\begin{equation}
\nabla_{\bf u}\hat{H}[{\bf u}](t) = 
(\hat{\bm{r}}\cdot \bm{\pi}) \nabla_{\bf u} \varepsilon[{\bf u}](t)
\end{equation}
Also note that a new auxiliary wave function $\chi[{\bf u}](t)$ has appeared; it is
defined as the solution of:
\begin{subequations}
\begin{equation}\label{eq-5a}
i\frac{\partial}{\partial t}\chi[{\bf u}](\bm{r},t) = 
\hat{H}^{\dag}[{\bf u}](t)\chi[{\bf u}](\bm{r},t)~\text{, and}
\end{equation}
\begin{equation}\label{eq-5b}
\chi[{\bf u}](\bm{r},T) = \hat{O}\Psi[{\bf u}]({\bf r},T)\,.
\end{equation}
\end{subequations}
These equations are similar to the equations for the true wave function
$\Psi[{\bf u}]$, except for the fact that the boundary condition
(Eq. \eqref{eq-5b}) is given at the final time $t=T$, which implies that
$\chi[{\bf u}]$ must be propagated backwards. Therefore, the computation of
the gradient of $G$, that requires of both wave functions, is computed by
first propagating Eq.~\eqref{eq-2} forward in time and then
Eq.~\eqref{eq-5a} backward.  Finally, the maxima of $G$ are found at the
critical points $\nabla_{\bf u}G=0$; in order to find these maxima we use
the quasi-Newton method designed by Broyden, Fletcher, Goldfarb and
Shanno~\cite{Fletcher2000}.

It remains to specify the target operator, and the parameterization of the
control functions. Regarding the former, we consider two types of targets:
\begin{description}
  \item[Target A:] The goal is to maximize the spin projection onto some
    direction $\bm{\xi}$, i.e. the operator $\hat{O}$ is defined as:
\begin{equation}\label{eq-7}
\hat{O} = \bm{\xi}\cdot\hat{\bm{\sigma}} =  \xi_x\hat{\sigma}_x+\xi_y\hat{\sigma}_y+\xi_z\hat{\sigma}_z\,,
\end{equation}
For example, if $\xi_z=1$, and $\xi_x=\xi_y=0$ the goal is to maximize the $z$ spin projection. 
The functional $F$ would therefore be defined as follows:
\begin{equation}\label{eq-9}
F\left[\Psi\right]=\langle\Psi(T)|\hat{O}|\Psi(T)\rangle=\sum_{i=x,y,z}\xi_i\langle\Psi(T)|\hat{\sigma}_i|\Psi(T)\rangle~\text{,}
\end{equation}
  \item[Target B:] The goal is to populate some selected excited state
    $\Phi_f$, that has the required spin orientation. The
    target operator is then defined as the projection onto that state:
\begin{equation}\label{eq-10}
\hat{O} = |\Phi_f\rangle\langle\Phi_f|\,,
\end{equation}
In this case, the functional $F$ is:
\begin{equation}\label{eq-11}
F\left[\Psi\right]=\langle\Psi(T)|\hat{O}|\Psi(T)\rangle=|\langle\Phi_f|\Psi(T)\rangle|^2\,.
\end{equation}
\end{description}

Finally, regarding the parametrization of the control function, we expand it
first in a Fourier series, and then we enforce several physical constraints:
The zero-frequency component is assumed to be zero (in order to ensure that
the signal over the full propagation time integrates to zero), and the sum of
all the cosine coefficients is also set to zero (in order to ensure that the
field will starts and ends at zero). In addition, we enforce a \emph{fixed
  fluence condition}:
\begin{equation}
\label{eq:fluence}
\int_0^T\!{\rm d}t\; \epsilon^2[{\bf u}](t) = F_0\,.
\end{equation}
The idea is to find the optimal field within the set of fields with equal
integrated intensity -- this is the physical meaning of the
fluence. The set of parameters ${\bf u}$ is constructed by considering first
the coefficients of the Fourier expansion, and them enforcing the mentioned
constraints -- for details, see Ref.~\cite{KKrieger2011}.

\section{Numerical results and analysis}
\subsection{Effect of Rashba SOC and magnetic field on the electronic structure}

\begin{figure}[t]
\includegraphics[width=15cm]{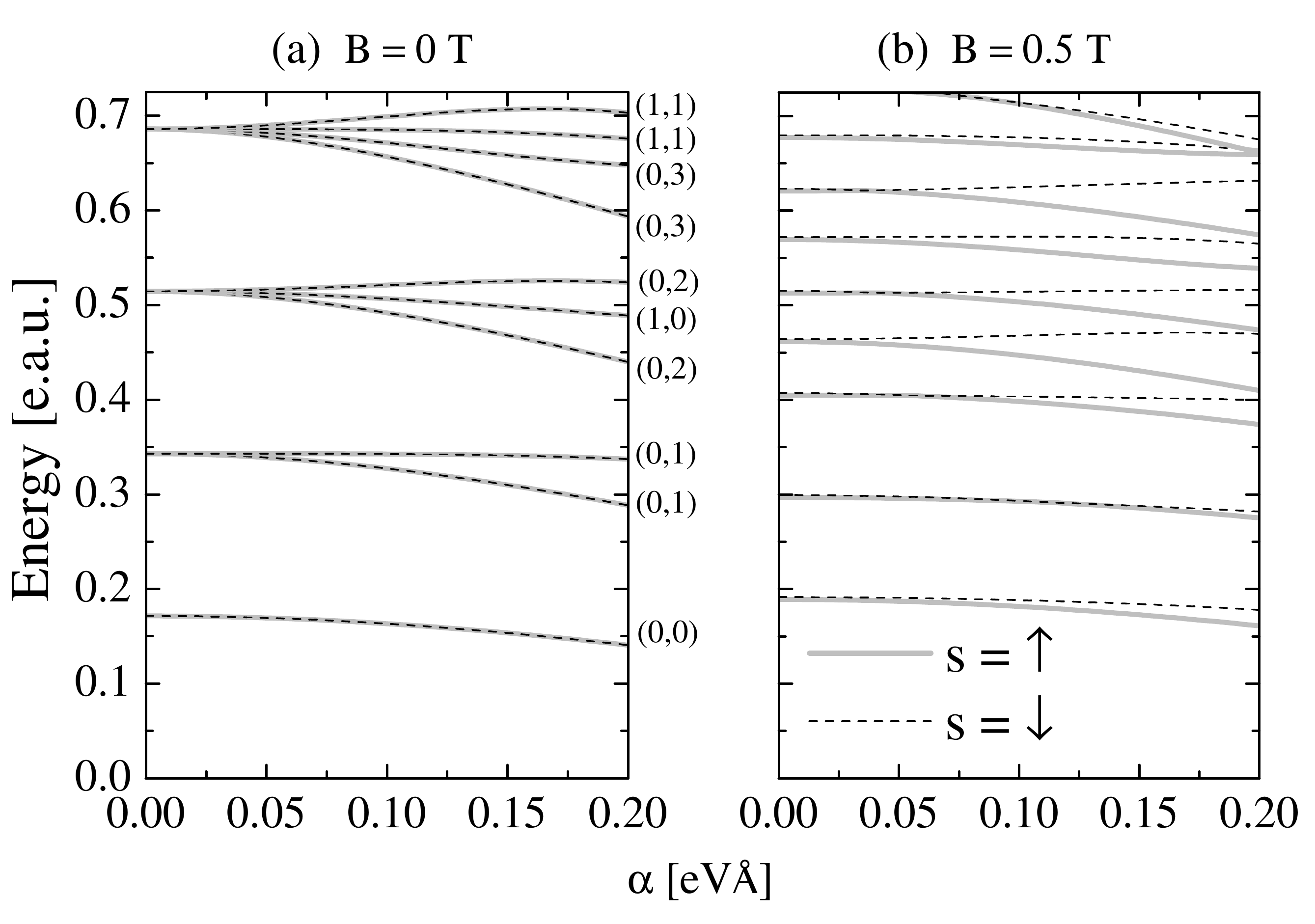}\\
\caption{Energy levels of the parabolic QD versus the strength of the Rashba SOC, $\alpha$, at (a) zero magnetic field and (b) $B=0.5$ T. We have labeled the energy levels in (a) as $(n,|l|)$.} \label{fig-1}
\end{figure}
\begin{figure}[t]
\includegraphics[width=15cm]{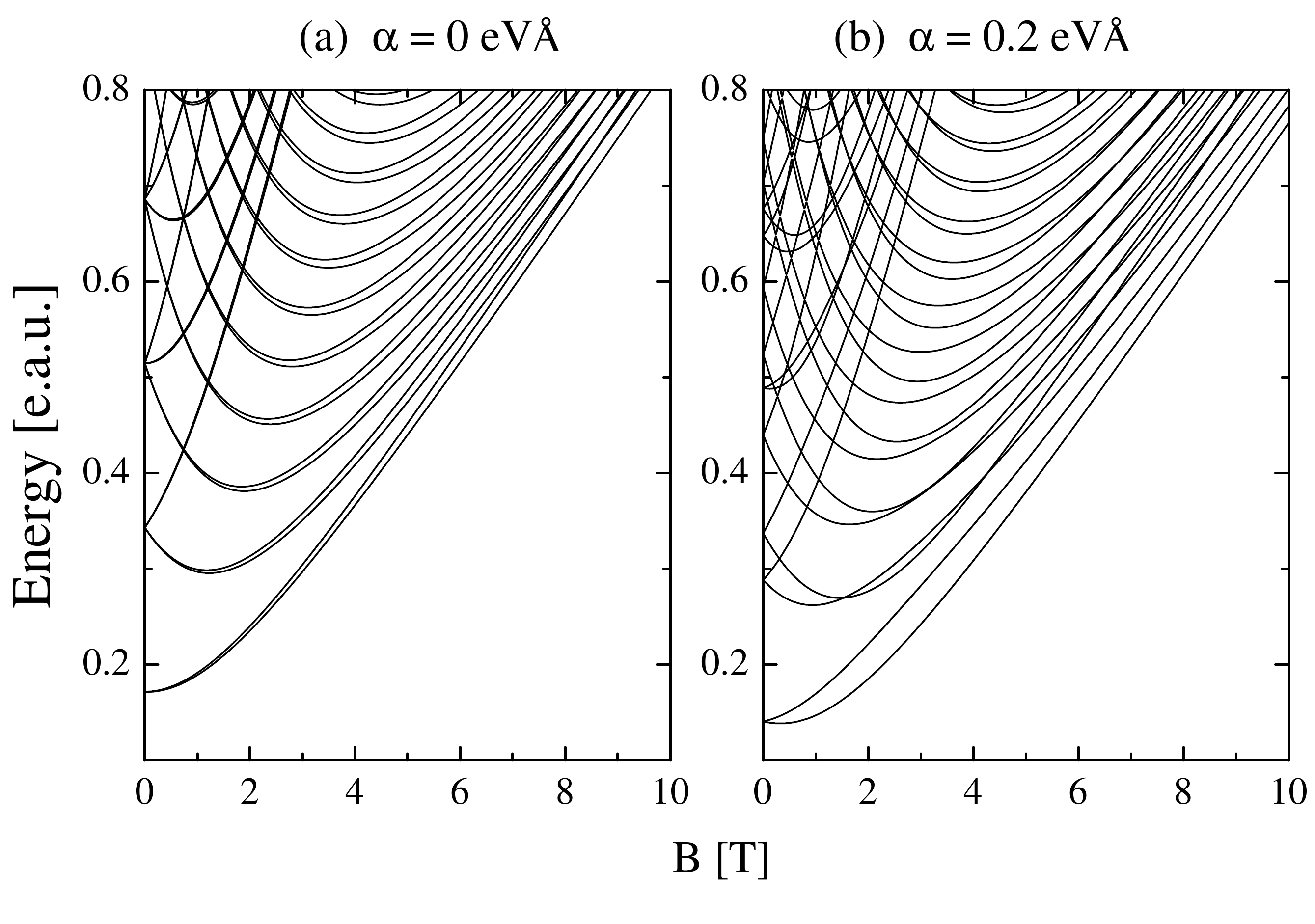}\\
\caption{Fock-Darwin spectrum of the parabolic QD with (a) $\alpha=0$ and (b) $\alpha=0.2$ eV{\AA}.} \label{fig-2}
\end{figure}

To start, we briefly review the effects of SOC on the eigenstates of the
QD. In the absence of this SOC, and of any magnetic field, the problem
determined by the Hamiltonian of Eq.~(\ref{eq-1}) is simply a 2D harmonic
oscillator. Therefore, the eigenstates are characterized by a principal
quantum number $n$, that spans a degenerate subspace of states that differ by
their orbital quantum number $l$, and their spin orientation, $s=\uparrow,\downarrow$.

In presence of the SOC, the picture changes: Fig.~\ref{fig-1} shows the QD
electronic energy levels as a function of $\alpha$ at (a) $B=0$ and (b)
$B=0.5$ T. In (a), the states are labeled by their $(n,l)$ numbers, which are
still good quantum numbers. In this zero magnetic field case, the electronic
levels undergo an energy shift due to the SOC. This displacement is
proportional to $\alpha^2$. The SOC also lifts the degeneracy of the
electronic states with the same orbital momenta (those with equal $|l|$). The
separation between these previously degenerate states is also proportional to
$\alpha^2$. Furthermore, at $B=0$, the energy levels remain doubly degenerate
in spin (Kramer's degeneracy).

When a magnetic field is present ($B=0.5$ T [Fig.~\ref{fig-1} (b)]), the usual Zeeman splitting appears. But in addition to this, we can observe the effect of the SOC: an $\alpha$-dependent spin splitting. As we increase the Rashba SOC strength, this splitting gives raise to an approach of energy levels with opposite magnetic moments. Note, however, that due to the SOC, the electronic states are no longer pure spin spin-up and spin-down states (in other words, $s$ is no more a good quantum number). Therefore, the labeling $s=\uparrow,\downarrow$ in Figs.~\ref{fig-1}(a) and~\ref{fig-1}(b) must be understood only in terms of the spin-branches of each electronic state.

Next, Fig.~\ref{fig-2} shows the Fock-Darwin spectra without Rashba SOC and with $\alpha=0.2$ eV{\AA}. It is clear in Fig.~\ref{fig-2}(b) that the SOC affects the dependence of the electronic states with the magnetic field in comparison with Fig.~\ref{fig-2}(a), lifting spin degeneracy even at vanishingly small magnetic field. In addition, new crossings of several of the energy levels at low magnetic field regime appears, as well as anti-crossings at higher magnetic field strengths. These anti-crossings occur between neighbouring quantum levels with opposite magnetic moments~\cite{PStano2005}. 

In the following sections, we will analyze the optimization
calculations. As described above, these optimizations are iterative
algorithms, and must depart from an initial electric pulse. In
all the cases discussed below, we start considering a ``reference'' pulse of the form:
\begin{equation}
\epsilon_{\textrm{ref}}(t)=\epsilon_0\cos{\left(\omega_0 t\right)}\cos{\left(\frac{\pi}{2}\frac{2t-T}{T}\right)}
\end{equation}
The peak amplitude $\epsilon_0$ is always set to $0.1$ kV$/$cm ($\sim
0.29$ e.a.u.). Nevertheless, in order to study the effect of initial
conditions on the optimization, on each case we have performed four
optimization runs starting from four different (random) initial laser
pulses, with the fluence [Eq.~(\ref{eq:fluence})] obtained from the
reference pulse and being preserved by the optimization procedure. The
results shown below correspond to the best outcomes.  The pulse
lengths, given by $T$, will be given in units of $\pi/\omega_0 \sim
1.15$ ps.  The pulses are then represented in a Fourier series, with
the constraints discussed above. One of them must obviously be the
establishemnt of a cut-off frequency.  For all the cases concerning
target type A, this cut-off frequency has been set to
$\omega_{\textrm{cut-off}}=20\left(2\pi/T\right)$, which implies 38
degrees of freedom (the number of parameters). For target B, we have
worked with $\omega_{\textrm{cut-off}}=10\left(2\pi/T\right)$ (18
degrees of freedom).

In the ground state, the expectation value of $\sigma_z$ is positive (if no
magnetic field is present, the ground state is two-fold degenerate in spin,
and then we choose the branch with positive $\langle\sigma_z\rangle$). The
goal that we want to achieve is to reverse this spin component. 
For that purpose, when using a target of type A, we set $\xi_z=-1$, and
$\xi_x=\xi_y=0$. For target B, we chose an eigenstate, $\Phi_f$, whose
$\langle\sigma_z\rangle$ component has opposite sign to that of the ground
state, $\Phi_i$.

\subsection{Target A: spin rotation at zero magnetic field}

\begin{figure}
\includegraphics[trim=0cm 0cm 0cm 0cm, clip=true,width=15cm]{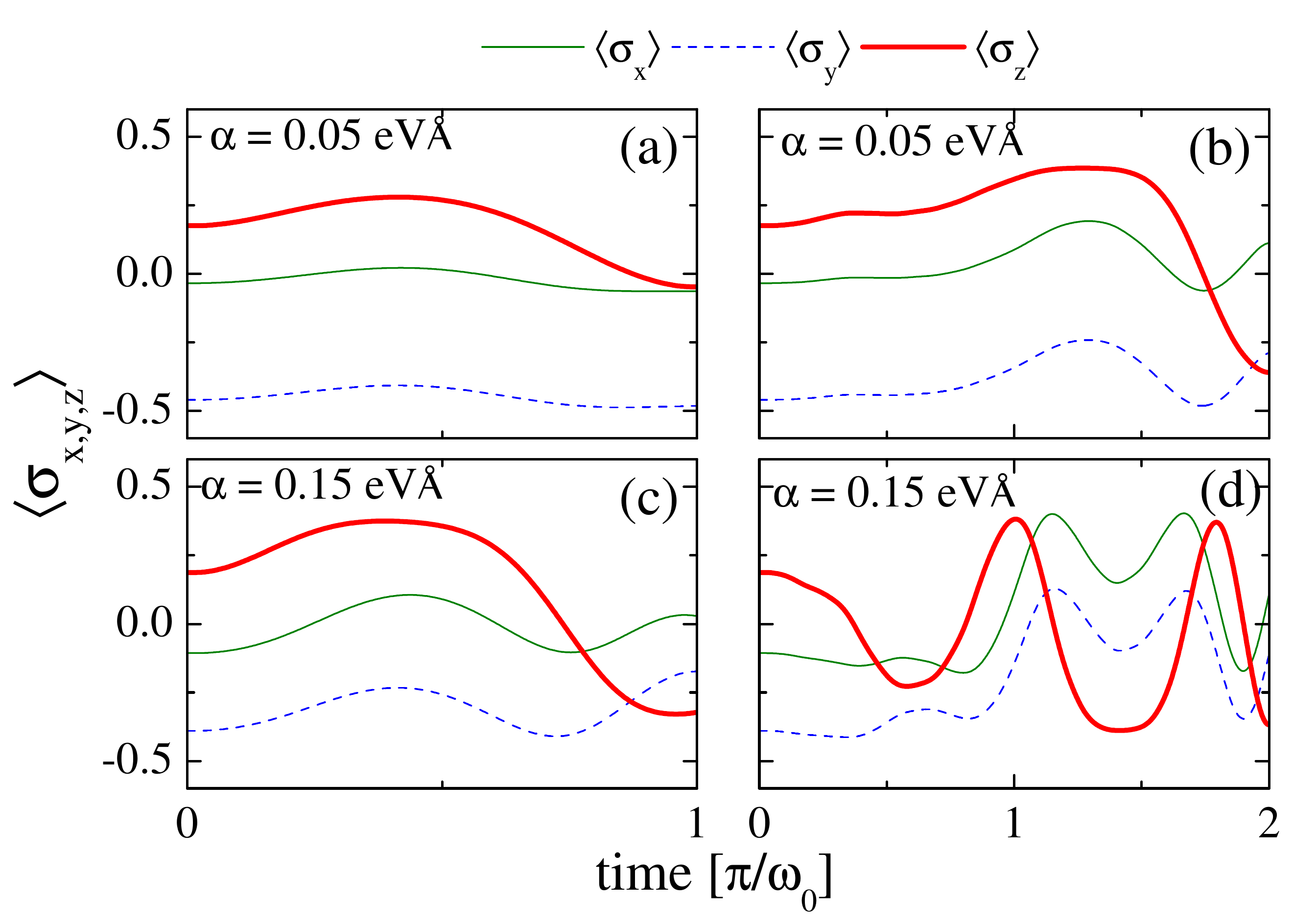}\\
\caption{(Color on-line) Average spin components (in units of $\hbar$) as a
  function of time for two optimized laser pulses of different lengths,
  $\pi/\omega_0\sim1.15$ ps and $2\pi/\omega_0\sim 2.3$ ps. We have considered
  two Rashba SOC strengths: $\alpha=0.05$ eV{\AA} (Figs. (a) and (b)) and
  $\alpha=0.15$ eV{\AA} (Figs. (c) and (d)).} \label{fig-3}
\end{figure}
\begin{figure}
\includegraphics[width=15cm]{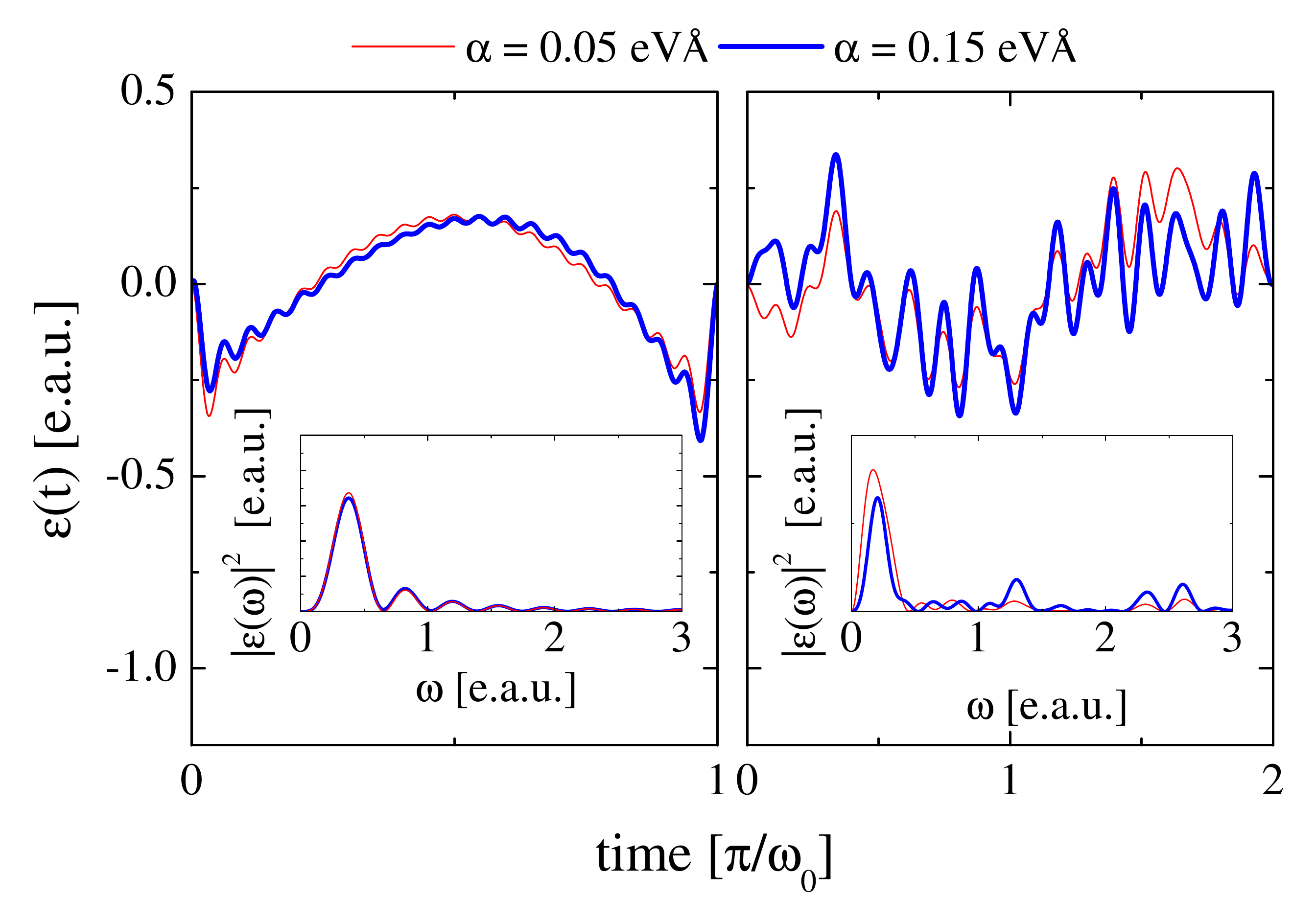}\\
\caption{(Color on-line) Optimal laser pulses corresponding to Fig.~\ref{fig-3}.} \label{fig-4}
\end{figure}

Fig.~\ref{fig-3} shows the time evolution of the average spin components for
two optimized laser pulse lengths: the left panels \ref{fig-3}(a) and
\ref{fig-3}(c) display shorter pulses ($T = \pi/\omega_0)$, and the right
panels correspond to double pulse lengths.  On the other hand, the top panels
\ref{fig-3}(a) and \ref{fig-3}(b) correspond to a weaker SOC strength,
$\alpha=0.05$ eV{\AA}, whereas the bottom panels \ref{fig-3}(c) and
\ref{fig-3}(d) correspond to a stronger $\alpha=0.15$ eV{\AA}. If we compare
the shorter pulses first ((a) and (c)) it becomes evident how the increased
strength of the Rashba SOC results in a spin orientation closer to the target
at the end of the pulse -- yet this optimization is still not significant.  An
increase in the pulse lengths, however, results in a very good final outcome
even for $\alpha=0.05$ eV{\AA}. In fact, as it can be seen in
Fig.~\ref{fig-3}(d), note that the component $\langle\sigma_z\rangle$
oscillates close to $-1/2$ even before the end of the pulse. This tells us
that, for that value of the SOC strength, an even shorter laser pulse duration
would suffice to reach a good $\langle \sigma_z \rangle$ value.

In Fig.~\ref{fig-4} we show the optimal laser pulses that produce the results
of Fig.~\ref{fig-3}. Note the very different aspect of the fields needed to
optimize the shorter and longer pulses -- the latter having a more complex
structure. We also display in the inset of each figure the power spectrum of
the pulses. The optimal pulses obtained with weaker and stronger SOC do not
differ significantly in shape, both in real time and in the frequency domain.

\subsection{Target A: spin rotation at non-zero magnetic field}

\begin{figure}
\includegraphics[width=10cm]{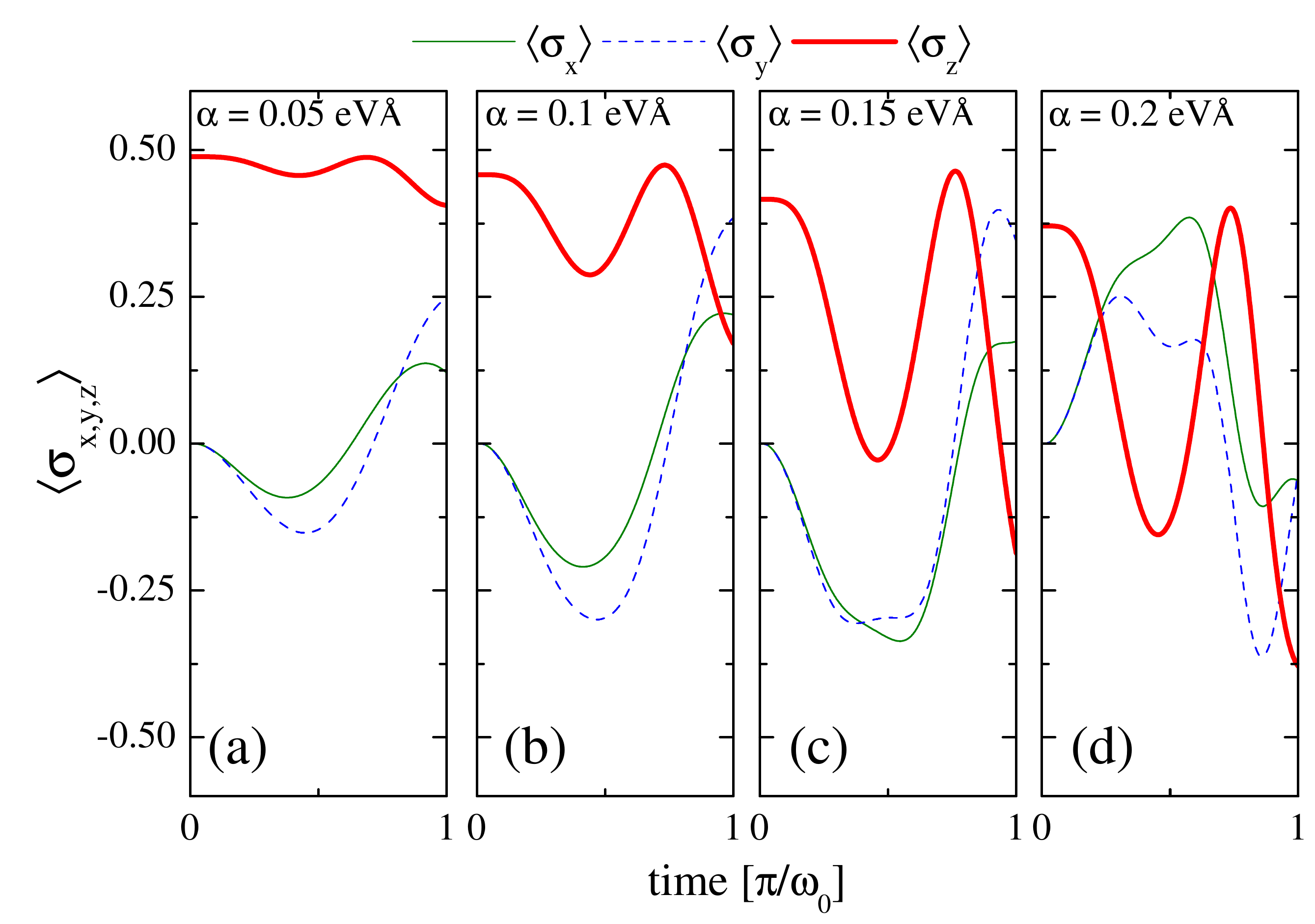}\\
\caption{(Color on-line) Average spin components as a function of time for an optimized laser pulse of length $\pi/\omega_0$, with $B=0.5$ T and four Rashba SOC strengths: (a) $\alpha=0.05$ eV{\AA}, (b) $\alpha=0.1$ eV{\AA}, (c) $\alpha=0.15$ eV{\AA} and (d) $\alpha=0.2$ eV{\AA}} \label{fig-5}
\end{figure}
\begin{figure}
\includegraphics[width=12cm]{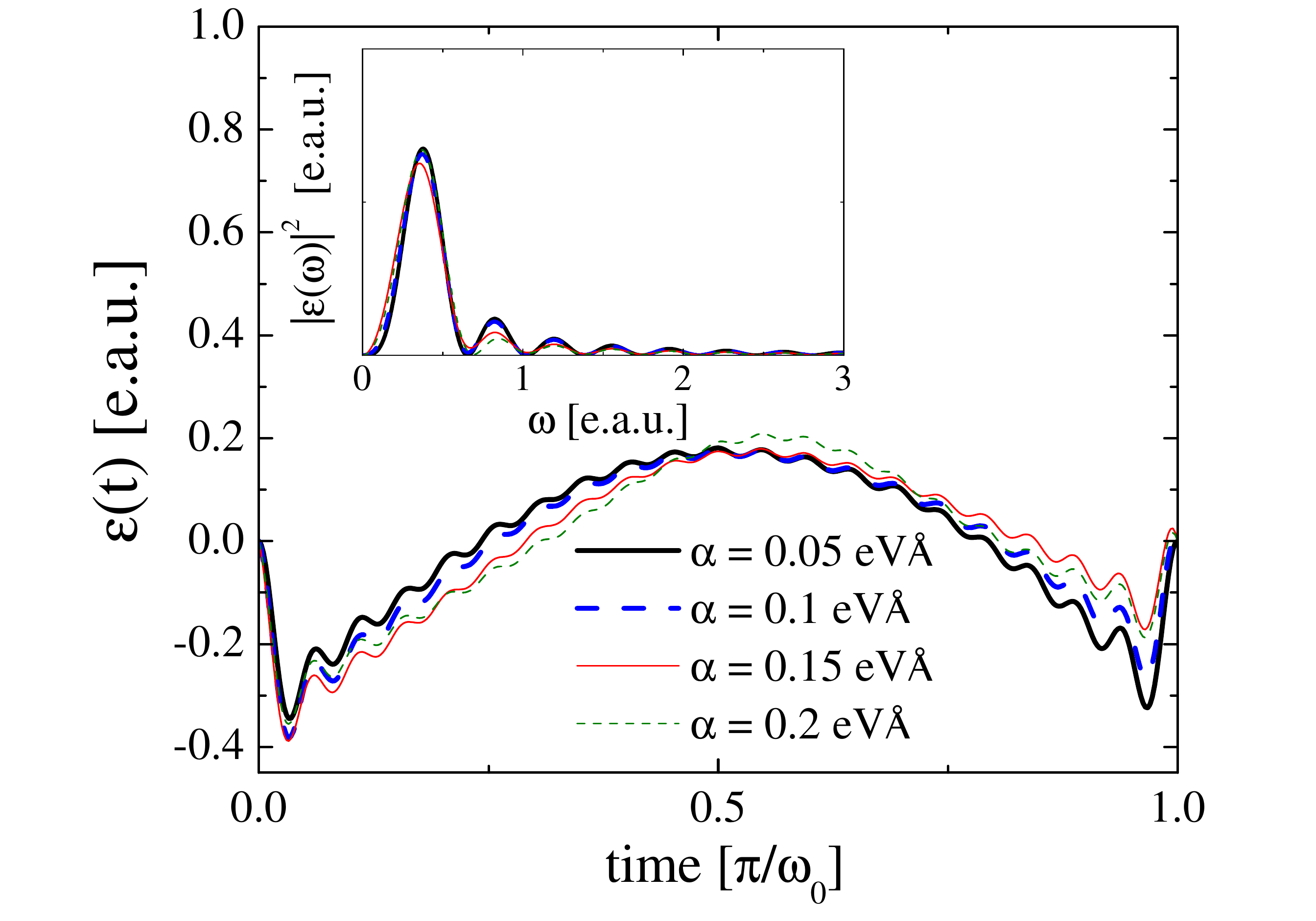}\\
\caption{(Color on-line) Optimized laser pulses corresponding to Fig.~\ref{fig-5}.} \label{fig-6}
\end{figure}

We now consider laser pulse optimizations in the presence of an external magnetic
field of $B=0.5$ T. Fig.~\ref{fig-5} shows the results for a pulse of length
$\pi/\omega_0$ and four different Rashba SOC strengths. The external magnetic
field competes now with the ``effective'' magnetic field associated to the
Rashba SOC and forces the ground state to have a $\langle\sigma_z\rangle$
closer to $1/2$. This is evident by looking at the value of
$\langle\sigma_z\rangle$ at the initial time in the figure,
compared with the initial values of the previous section. We can also observe
that increasing the Rashba SOC allows to improve the spin rotation, getting a
result very close to the target for $\alpha=0.20$ eV{\AA}. Note that, due to
the presence of the magnetic field, the starting point is farther from the
target, compared to the cases with zero magnetic field, which means that the
spin state must perform a longer path. Fig.~\ref{fig-6} shows the optimal
pulses associated to Fig.~\ref{fig-5}. They are qualitatively similar to those
shown in Fig.~\ref{fig-4}.

\begin{figure}[t]
\includegraphics[width=10cm]{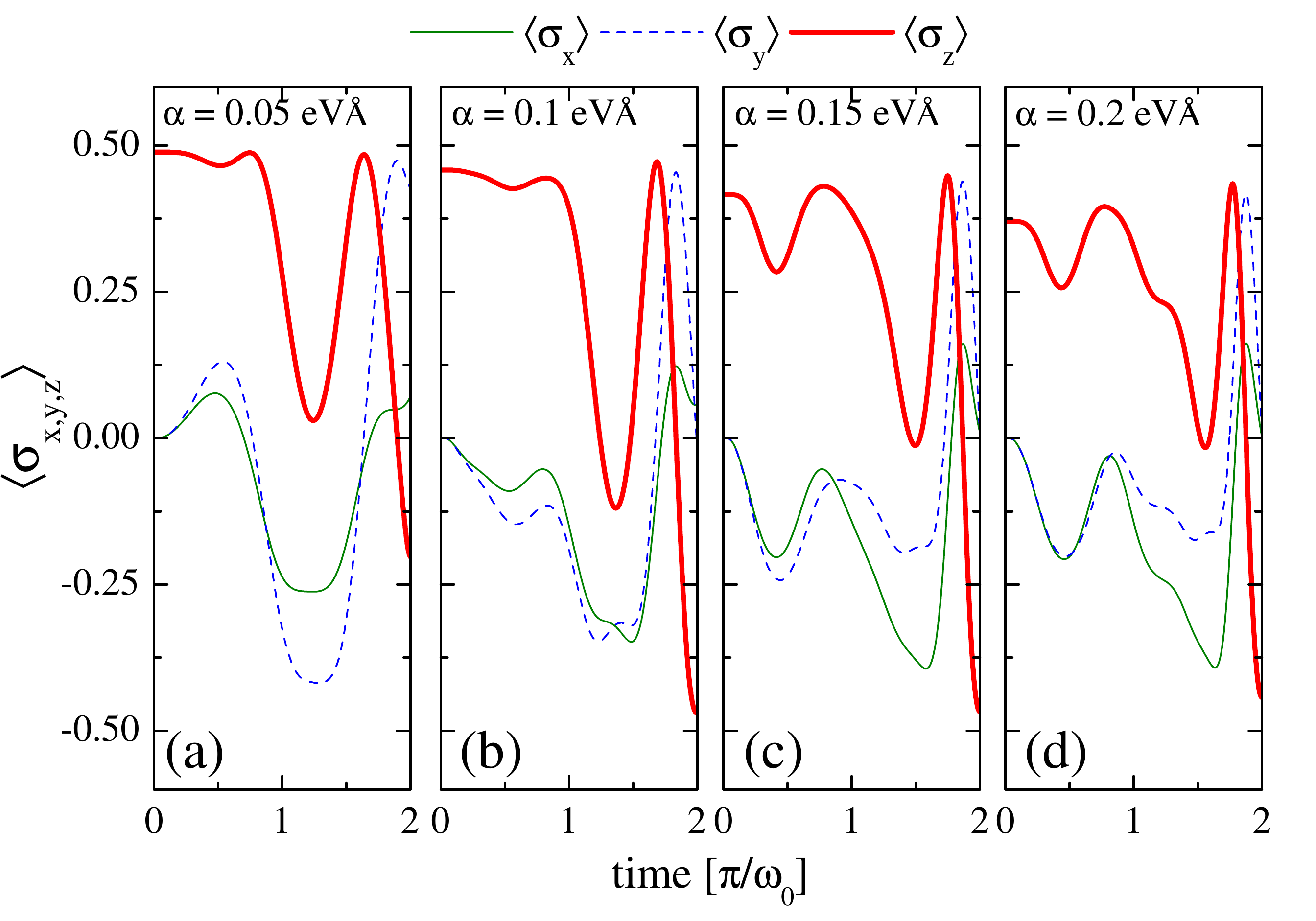}\\
\caption{
(Color on-line) Average spin components as a function of time for an optimized
laser pulse of length $\pi/\omega_0$, with $B=0.5$ T and four Rashba SOC
strengths: (a) $\alpha=0.05$ eV{\AA}, (b) $\alpha=0.1$ eV{\AA}, (c)
$\alpha=0.15$ eV{\AA} and (d) $\alpha=0.2$ eV{\AA}.} 
\label{fig-7}
\end{figure}

\begin{figure}[t]
\includegraphics[width=12cm]{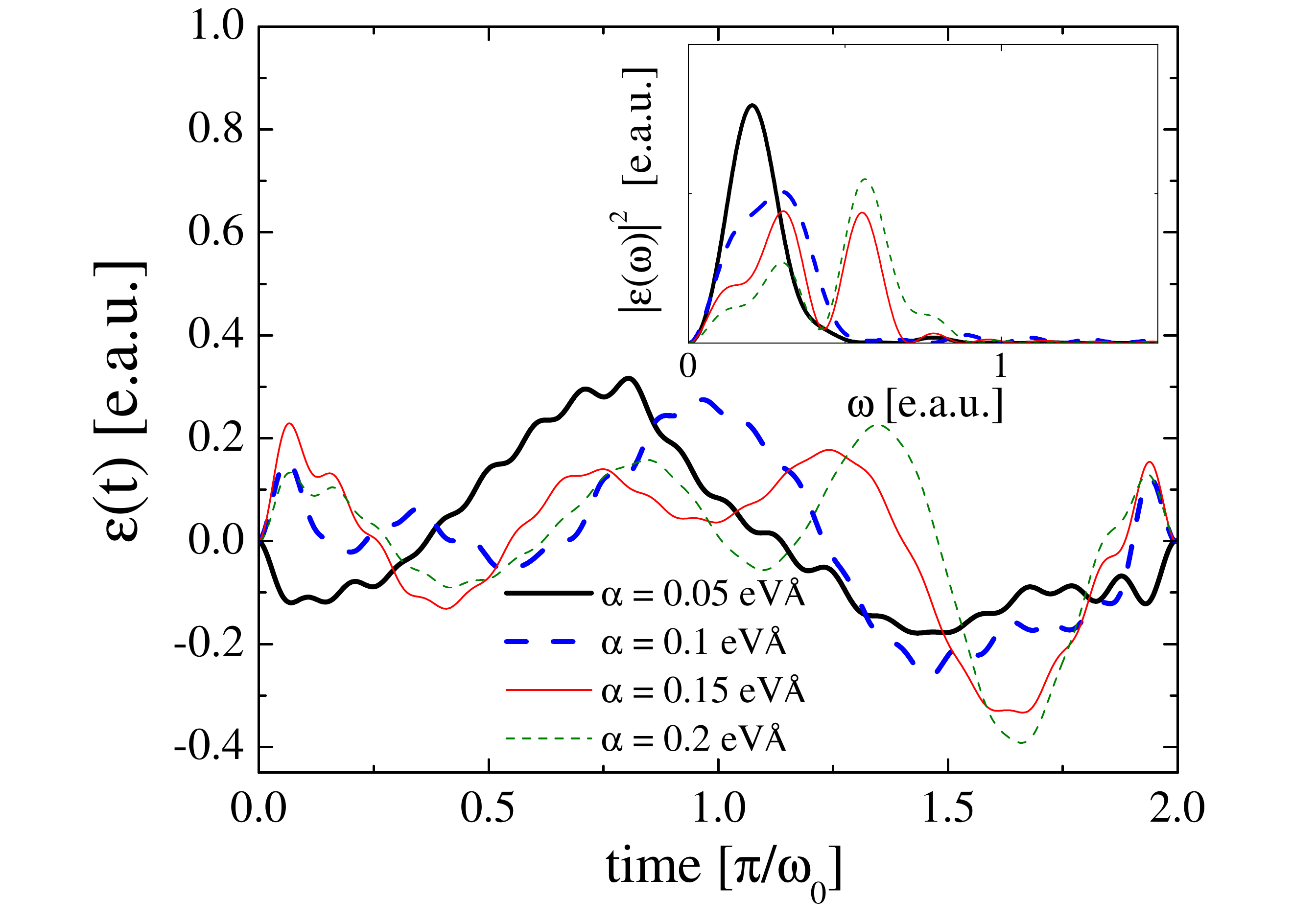}\\
\caption{(Color on-line) Optimized laser pulses corresponding to Fig.~\ref{fig-7}.} \label{fig-8}
\end{figure}

Let us now consider pulse lengths of $2\pi/\omega_0$ (Fig.~\ref{fig-7}). In
this case, notice that the optimization algorithm is capable of finding pulses
that reach the target with smaller SOC strengths than those required with
shorter pulses. Regarding the optimal pulses (Fig.~\ref{fig-8}), note that
these look very different to the shorter pulses of length $\pi/\omega_0$. One
can see how the increase of the SOC strength produces an
apparent increase in the number of oscillations of the pulse envelope. In the
inset of this figure, note how this increase of the SOC strength results in a
splitting of the initial single band into two separate bands. The
lower frequency band is associated with the oscillations of the envelope and
moves to higher frequency than the initial band as we increase the strength of
the SOC. 

\begin{figure}[t]
\includegraphics[width=12cm]{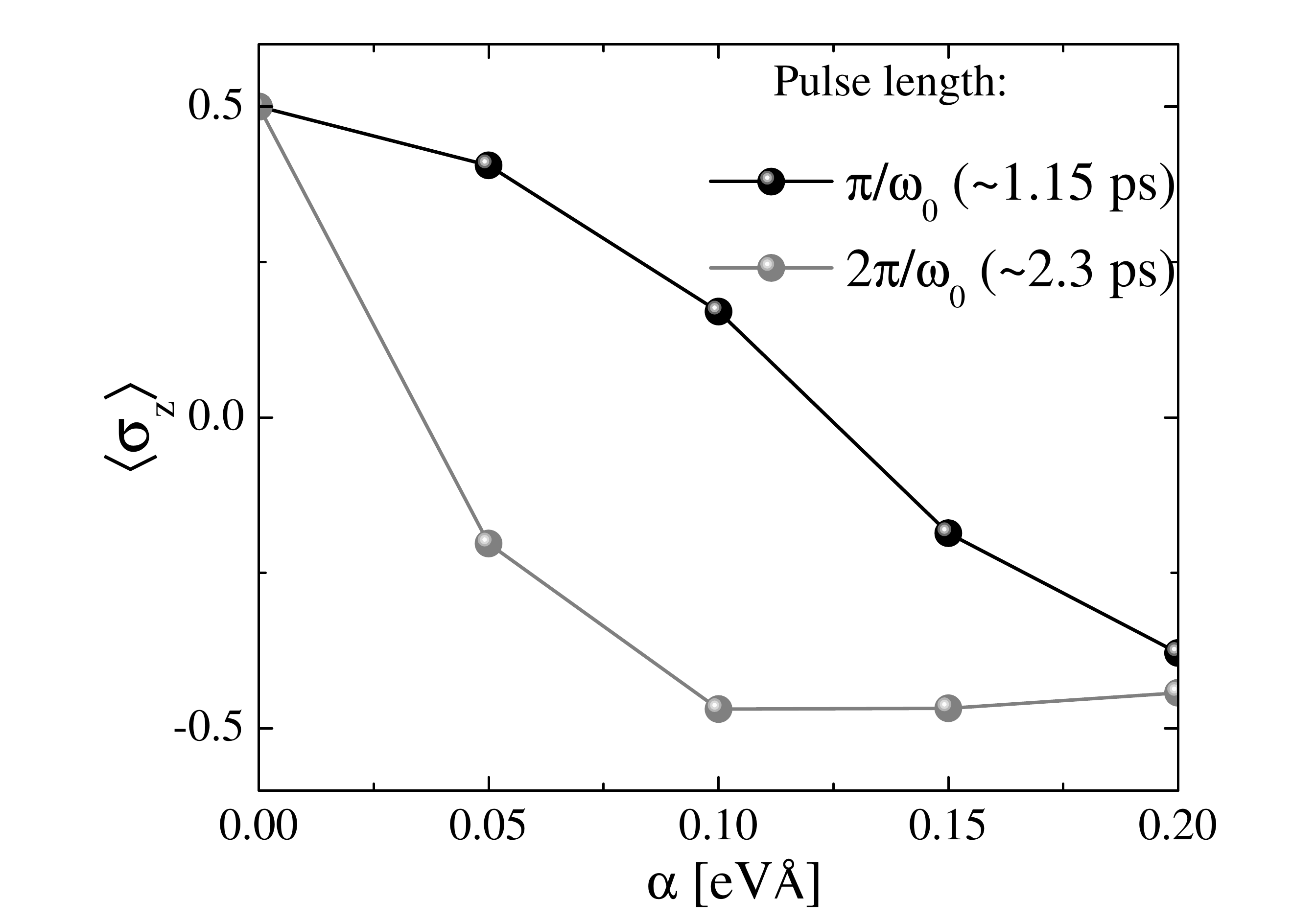}\\
\caption{Expectation value of the $z$-component of the spin at $t=T$ as a function of Rashba SOC strength for two laser pulse lengths.} \label{fig-9}
\end{figure}

Finally, Fig.~\ref{fig-9} shows the expectation value of $\sigma_z$, at the
end of the pulse, as a function of $\alpha$ for the two pulse lengths
considered. Here it becomes evident how the increase of pulse length allows
for a better result. It is also evident that one may get a faster spin-flip
through the tuning (increase) of the Rashba SOC.


\subsection{Target B: spin rotation at non-zero magnetic field}

We now describe the results obtained when using a target of type B. We have
chosen to maximize the transition between the two spin branches of the ground
state, i.e., $E_{0,0,\uparrow} \rightarrow E_{0,0,\downarrow}$ (in the
following, the figures will label these two states as $i = 0$ and $i = 1$,
respectively). This transition is only allowed when the two branches are split
by the effect of a magnetic field, and therefore, for this calculation, we have set the magnetic field and Rashba SOC strengths as $0.5$ T and
$0.2$ eV{\AA}, respectively. The cutoff frequency, set in this case as
$10\left(2\pi/\omega_0\right)$, is well above the resonant frequency
associated with this transition, $\omega_{0\rightarrow1}$.

\begin{figure}[t]
\includegraphics[width=12cm]{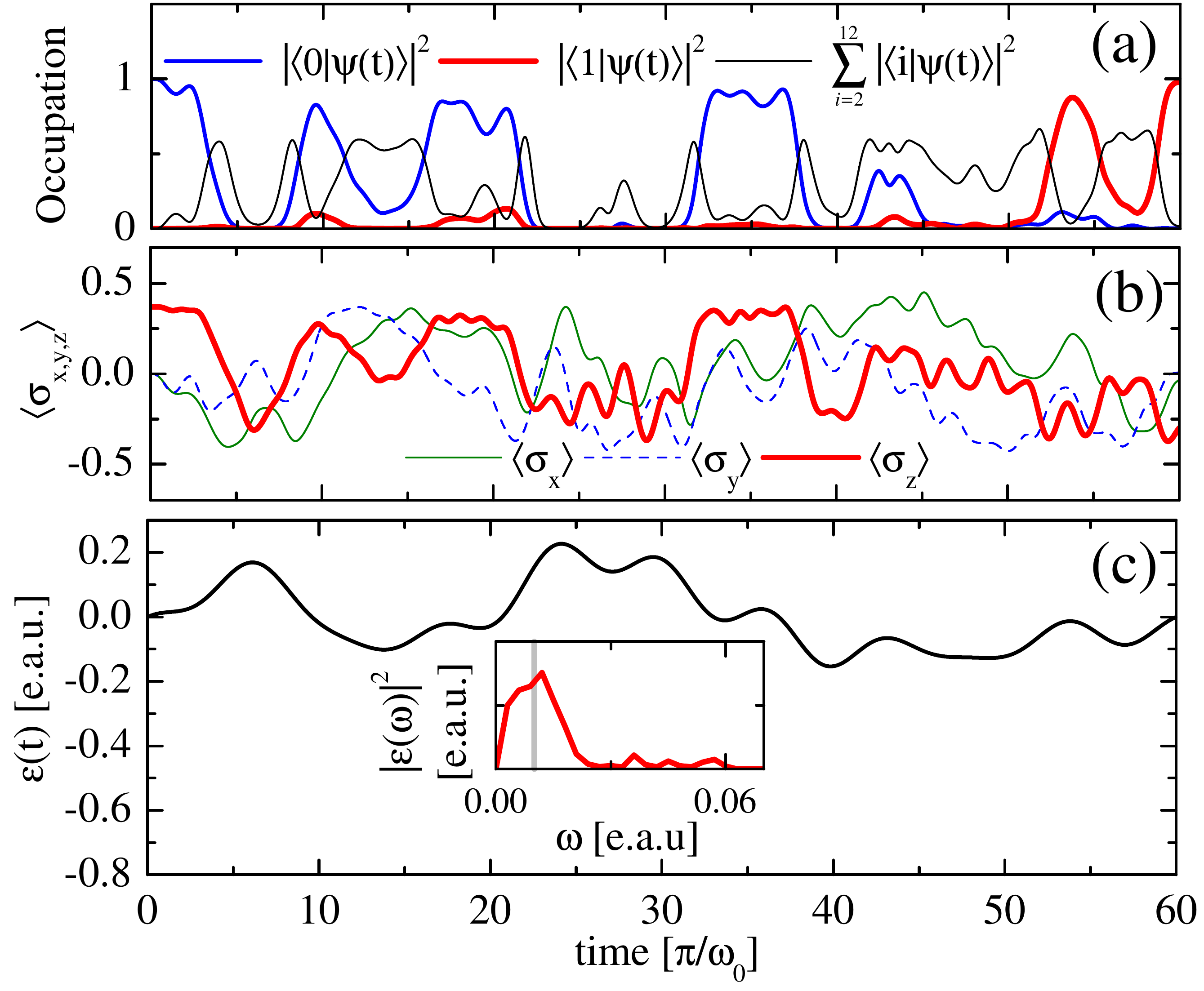}\\
\caption{
(Color on-line) (a) Occupation of the 12 first eigenstates as a function of
time, (b) average spin components as a function of time and (c) Optimized laser pulse. The length of the pulse is $60\left(\pi/\omega_0\right)$, $B=0.5$ T
and $\alpha=0.2$ eV{\AA}. The inset in (c) shows the power spectrum in frequency of the optimized laser pulse.} 
\label{fig-10}
\end{figure}

Fig.~\ref{fig-10}(a) shows the time evolution of the occupation of the first
$12$ eigenstates of the QD in response to an optimized laser field of length
$60\left(\pi/\omega_0\right)$. The occupation of state $i=1$
reaches a maximum occupancy of $\sim 0.98$ at the end of the pulse, and the
occupation of the ground state decays to zero quickly. During the course of
the pulse, the global occupation of those 12 first eigenstates decays almost to
zero several times, indicating that in these time intervals the electron is
occupying higher energy states. Below, Fig.~\ref{fig-10}(b) shows
the time evolution of $\langle\sigma_x\rangle$, $\langle\sigma_y\rangle$ and
$\langle\sigma_z\rangle$. As expected, the final spin
orientation of the electron agrees almost completely with the expectation
values of the spin components in state $i=1$. 

Fig.~\ref{fig-10}(c) shows the associated optimal laser field. It is important
to remember that, in the case of target B, we attempt a state-to-state
transition of frequency $\omega_{0\rightarrow1}$, indicated by a vertical
thick gray line in the inset of Fig.~\ref{fig-10}(c). In general, despite the
random character of the initial guess field of the search, the optimal
pulse is characterized by the presence of a wide distribution of frequencies
around $\omega_{0\rightarrow1}$.

This is of course not surprising. However, the optimal pulse is not merely a
quasi mono-chromatic pulse with the transition frequency. We have in fact
attempted the optimization by starting from these type of pulses, finding that
the optimization alters that starting point by adding the necessary extra
frequencies to obtain a significantly better result. These runs (not shown
here), provided better solutions than the runs started from purely random
pulses. Finally, we note that the complexity of the transition process found
by the QOCT procedure is also evidenced from the population of many
eigenstates during the evolution, far from the two-state model that is used to
explain Rabi oscillations. This population of ``auxiliary'' states is of
course also present when using target A, a fact to be expected since the
states in these case are not ``controlled''. Indeed, when using target A even
the final state is composed of a superposition of a large number of
eigenstates.

\section{Conclusions}\label{conclusions}      

We have theoretically demonstrated the possibility of manipulating
the electronic spin in a semiconductor QD, by means of ultrashort laser
pulses, making use of the spin-orbit coupling. We have explored the time
scales necessary to perform spin transformations, making use of optimized
laser pulse shapes, found with the help of quantum optimal control theory.
These time scales depend on the strength of the spin-orbit coupling, and of
the presence or absence of an external magnetic field, helpful to fix a value
for the initial spin orientation. 

The search for an optimal pulse, within QOCT, can be done in multiple ways,
and the first choice to make is the design of a target functional.  We have
shown results for two types of targets. In the first type, the functional only
depends on the spin projection value, without placing explicit restrictions on
the number of the QD eigenstates which can participate in the representation
of the final state. The results show full spin rotations in about 1 or 2 ps,
depending on the value of the spin-orbit coupling. In the second type of
target, the spin control is achieved indirectly through the control of a
particular transition between the ground state and an excited state which has
the desired spin orientation. In this case, if the target transition involves
one of the lowest excited states, the pulse length must be at least an order
of magnitude greater than those used in the case of the first target.

In conclusion, our simulations support the idea of ultrafast manipulations of
electronic spin in 2D quantum dots, by means of the electric fields of THz
laser pulses, using spin-orbit coupling to transform the electric signal into
a spin rotation.

\section{Acknowledgements}\nonumber

This work was supported by the European Community's FP7 through the CRONOS
project, grant agreement no. 280879.

\widetext

\end{document}